\documentclass[%
reprint,
superscriptaddress,r
 amsmath,amssymb,
 aps,
pra,
]{revtex4-1}

\usepackage{graphicx}
\usepackage{dcolumn}
\usepackage{bm}
\usepackage{amsmath}
\usepackage{xcolor}

\begin{document}


\title{Engineering atomic superradiance scaling in cavity QED system with collective and individual emission channels}

\author{Ruijin Sun}
\affiliation{%
 Center for Quantum Science and School of Physics, Northeast Normal University, Changchun 130024,China
}%
\affiliation{%
 School of Physics, University College Cork, Cork, Munster T12 K8AF, Ireland}%
\author{Xiang Guo}
\affiliation{%
 Center for Quantum Science and School of Physics, Northeast Normal University, Changchun 130024,China
}%
\author{Andreas Ruschhaupt}
\affiliation{%
School of Physics, University College Cork, Cork, Munster T12 K8AF, Ireland}%
\author{Zhihai Wang}
\email{wangzh761@nenu.edu.cn}
\affiliation{%
 Center for Quantum Science and School of Physics, Northeast Normal University, Changchun 130024,China
}%


\begin{abstract}

The coherent emission of multiple atoms gives rise to superradiance, a cornerstone phenomenon in quantum optics with wide-ranging applications in quantum information processing and precision metrology. Despite its importance, how the superradiant scaling with respect to the number of participating atoms can be effectively controlled remains largely unexplored. In this work, we investigate a cavity-QED system and demonstrate that  atom-photon coupling can significantly alter the emission behavior--suppressing the collective superradiant scaling while enhancing the scaling associated with individual atomic emissions. Our study provides a pathway toward controllable collective emission in state-of-the-art experimental platforms.

\end{abstract}

\maketitle


\section{Introduction}

Dicke superradiance~\cite{PhysRev.93.99, zhang2025unraveling,
andreev1980collective,mlynek2014observation,wang2020supercorrelated,zhu2024single,guo2024phonon, gross1982superradiance} describes the phenomenon where an ensemble of $N$ excited atoms undergoes collective spontaneous emission, leading to an enhanced radiation strength that scales as $N^2$, in contrast to the linear $N$ scaling for independent emission. This cooperative effect arises from the constructive interference among the atomic dipoles and has found wide applications in lasers, quantum metrology, and precision measurements~\cite{bohnet2012steady, ferioli2021laser, paulisch2019quantum,koppenhofer2023squeezed}. In recent years, superradiance has garnered significant interest in cavity and waveguide quantum electrodynamics (QED) platforms~\cite{sheremet2023waveguide, cardenas2023many, windt2025effects}, where natural or artificial atoms act as emitters, and the confined photonic modes provide a shared reservoir that mediates and enhances the interference effects.

The strong coupling between light and matter in cavity quantum electrodynamics (QED) enables the observation of quintessential quantum phenomena such as Rabi oscillations and vacuum Rabi splitting~\cite{mukhopadhyay2024quantum,PhysRevLett.110.066802}, photon blockade~\cite{trivedi2019photon,hamsen2017two}, and the generation of nonclassical light~\cite{mlynek2014observation,majumdar2012probing}. With rapid advances in experimental techniques, the coherent interaction between a single two-level atom and a single-mode cavity field~\cite{kimble1998strong,ye1999trapping,boozer2006cooling,thompson2013coupling} has been extended to systems involving multiple atoms~\cite{colombe2007strong,liu2023realization}, where collective radiative effects emerge. Despite these developments, a systematic understanding of how strong atom–photon coupling modifies the collective emission scaling in multi-atom systems remains elusive.

In contrast, when an array of $N$ excited atoms independently interacts with its local environment, interference among the atomic emission channels vanishes, leading to radiation strength that scales linearly with $N$, without superradiant enhancement. Recently, experiments have realized the coupling of an array of tens of two-level atoms to a single cavity mode and observed a clear atom-number-dependent Rabi splitting~\cite{yan2023superradiant}. Nevertheless, how the cavity functions as a quantum data bus that reshapes their emission dynamics has not yet been thoroughly explored.

To fill these gaps and clarify how the cavity controls atomic radiance, we systematically examine both collective and individual atomic emission processes in a cavity-QED configuration, where the atoms form an ensemble or a spatially separated array, respectively. To overcome the exponential wall difficulty in straightforward brute-force solution of the master equation, we adopt semiclassical phase-space methods, namely, the truncated Wigner approximation (TWA) and the discrete truncated Wigner approximation (DTWA)~\cite{huber2021phase,mink2023collective,huber2022realistic,
khasseh2020discrete,hao2021observation}--to study collective and individual atomic emission, respectively. These approaches incorporate part of the quantum fluctuations by averaging over many stochastic trajectories and introducing random noise to each coarse-grained moment. As such, they go beyond the commonly used mean-field approximation~\cite{opper2001advanced}. Moreover, this classical treatment reduces the computational complexity from exponential to linear depending on the number of atoms. As a result, the dynamics of systems involving hundreds or even thousands of atoms becomes computationally accessible.

For collective emission, we find that the characteristic $N^2$ superradiant scaling is suppressed when the atoms are strongly coupled to an auxiliary leaky cavity mode. In contrast, in the individual emission configuration, the atom-photon coupling enhances the emission scaling from the conventional linear $N$ behavior toward $N^{1.9}$, approximate quadratic scaling characteristic of Dicke superradiance. These results demonstrate a viable approach to control and engineer superradiant scaling in cavity QED systems.

This paper is organized as follows. In Sec.~\ref{s2}, we investigate the superradiance scaling in a cavity QED setup, where the atoms emit collectively, using the TWA approach. In Sec.~\ref{s3}, we analyze the scaling behavior for the case of individual atomic emission with the aid of the DTWA method. In Sec.~\ref{s4}, we  present a brief conclusion.

\section{Collective atomic emission}
\label{s2}
\subsection{Model}

\begin{figure}
\begin{center}
\includegraphics[width=1 \columnwidth]{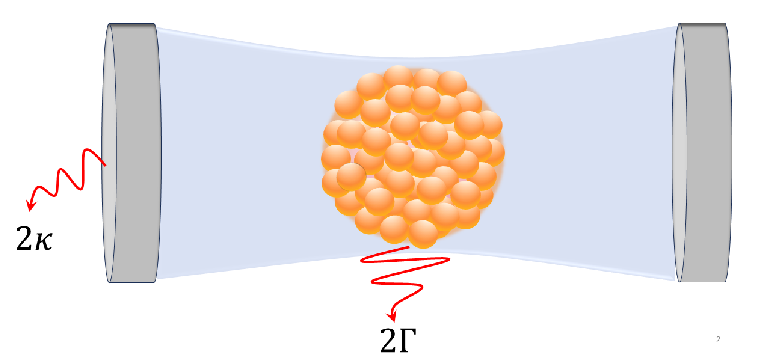}
\end{center}
\caption{\label{FIG.1}Schematic illustration of the collective emission process in a multi-atom cavity QED system. The cavity mode decays at a rate $2\kappa$, and the atomic ensemble exhibits collective radiation with an effective decay rate $2\Gamma$.}
\label{scheme1}
\end{figure}

In this section, we consider an ensemble of $N$ identical two-level atoms interacting with a single-mode cavity field, as illustrated in Fig.~\ref{scheme1}. Neglecting direct atom-atom interactions, the Hamiltonian of this system under the rotating wave approximation reads (here and after, we set $\hbar=1$)
\begin{equation}
H = \frac{\omega_{a}}{2} S_{z} + \omega_{c} c^{\dagger}c + g(S^{+}c + S^{-} c^{\dagger}),
\label{commonH}
\end{equation}
where $\omega_{a}$ denotes the atomic transition frequency, $S_{z} = \sum_{i=1}^{N} \sigma_{z}^{i} / 2$ is the total Pauli-$z$ operator, and $S^{+} = \sum_{i=1}^{N} \sigma_{+}^{i}$ and $S^{-} = \sum_{i=1}^{N} \sigma_{-}^{i}$ are the collective atomic raising and lowering operators, respectively.  $\sigma^{i}_{z}$ and $\sigma^{i}_{\pm}$ are the Pauli-$z$ and raising/lowering operators acting on the $i$-th ($i=1,2,\cdots ,N$) two-level atom. $c(c^{\dagger})$ is the annihilation (creation) operator of the single-mode cavity field with resonance frequency $\omega_{c}$. $g$ is the atom-cavity coupling strength.

With cavity decay and atomic spontaneous emission considered, the dynamics of the system is governed by the master equation
\begin{eqnarray}
\partial_{t} \rho &=& -i [H, \rho]
+ 2\kappa ( c \rho c^{\dagger} - \frac{1}{2} \{ c^{\dagger} c, \rho\} ) \nonumber \\
&& + 2\Gamma ( S^{-} \rho S^{+} - \frac{1}{2} \{S^{+} S^{-}, \rho\} ),
\label{masterq}
\end{eqnarray}
where $2\kappa$ and $2\Gamma$ denote the decay rates of the cavity mode and the atoms, respectively, and $\{A,B\}=AB+BA$.

In principle, the master equation can be solved numerically. However, the dimension of the Liouvillian space scales as $[(N+1)\times N_a]^2$, where $N_a$ is the photon number cutoff for the cavity mode. As the number of the atoms increases, a larger photon cutoff $N_a$ is needed to capture the full dynamics, which further enlarges the Liouvillian dimension. Therefore, brute-force numerical simulations with the master equation are in practice not achievable for very large system.

To overcome this challenge, we employ the TWA to simulate the system dynamics. Within this framework, we use the Schwinger boson representation to map the collective spin operators to bosonic modes ~\cite{huber2021phase,olsen2005phase,ng2011exact}:
\begin{equation}
S^{+} = a^{\dagger} b,\quad
S^{-} = a b^{\dagger},\quad
S_{z} = \frac{1}{2}(a^{\dagger} a - b^{\dagger} b),
\end{equation}
where $a$ and $b$ are bosonic annihilation operators. This transformation enables a tractable semi-classical description of the collective atomic degrees of freedom in phase space.

As the next step, we introduce the parameterized Wigner phase-space distribution function
\begin{equation}
F(\vec{x},t)=\frac{1}{\pi^{4N}}\int d^{4N} \lambda e^{(\vec{x}\vec{\lambda}^{*}-\vec{x}^{*}\vec{\lambda})} \mathrm{Tr} \{e^{\vec{\lambda}\vec{v}^{\dagger}} \rho (t) e^{-\vec{\lambda}^{*}\vec{v}} \} e^{\frac{|\vec{\lambda} |^{2}}{2}},
\end{equation}
where $\vec{v}=(a,b,c)$ is the vector of bosonic annihilation operators, $\vec{x}$ and $\vec{\lambda} $ are vectors containing the same number of complex variables corresponding to these operators ($a,b,c\rightarrow \alpha,\beta,\eta$). We then apply the single-mode cavity field mapping $c(c^\dagger)$ from the master equation to the Fokker-Planck equation as~\cite{huber2022realistic}:
\begin{subequations}
\begin{align}
c\rho &\rightarrow \left[ \eta + \frac{1}{2} \frac{\partial}{\partial \eta^{*}} \right] F, \quad
c^{\dagger}\rho \rightarrow \left[ \eta^{*} - \frac{1}{2} \frac{\partial}{\partial \eta} \right] F,  \\
\rho c^{\dagger} &\rightarrow \left[ \eta^{*} + \frac{1}{2} \frac{\partial}{\partial \eta} \right] F, \quad
\rho c \rightarrow \left[ \eta - \frac{1}{2} \frac{\partial}{\partial \eta^{*}} \right] F,
\end{align}
\end{subequations}

As a result, the master equation, Eq.~\eqref{masterq}, is transformed into the Fokker -Planck equation
\begin{align}
\frac{\partial}{\partial t} F = &
\left[ \frac{\partial}{\partial \alpha} \left( i\frac{\omega_a}{4} \alpha + ig \beta \eta + \Gamma\left( |\beta|^2 + \frac{1}{2} \right)\alpha \right) \right. \nonumber \\
& + \frac{\partial}{\partial \beta} \left( -i \frac{\omega_a}{4} \beta + ig \alpha \eta^* - \Gamma\left( |\alpha|^2 - \frac{1}{2} \right) \beta \right) \nonumber \\
& + \left. \frac{\partial}{\partial \eta} \left( ig \alpha \beta^* + i\omega_c \eta + \kappa \eta \right) + \mathrm{c.c.} \right] F \nonumber \\
& + \left[
\frac{\partial^2}{\partial \alpha \partial \alpha^*} \Gamma\left( |\beta|^2 + \frac{1}{2} \right)
+ \frac{\partial^2}{\partial \beta \partial \beta^*} \Gamma\left( |\alpha|^2 - \frac{1}{2} \right) \right. \nonumber \\
& + \frac{\partial^2}{\partial \eta \partial \eta^*} \kappa
- \frac{\partial^2}{\partial \alpha \partial \beta} \Gamma \alpha \beta
- \left. \frac{\partial^2}{\partial \alpha^* \partial \beta^*} \Gamma \alpha^* \beta^* \right] F.
\label{FP}
\end{align}

Eq.~(\ref{FP})  can be written as a partial differential equation for the phase-space distribution $F$ with drift vector $\vec{A}$ and diffusion matrix $D$:
\begin{equation}
    \frac{\partial}{\partial t}F(\vec{x},t) = \left[-\sum_{j} \frac{\partial}{\partial x_j} A_j(\vec{x})
    + \frac{1}{2} \sum_{i,j} \frac{\partial^2}{\partial x_i \partial x_j^*} D_{ij}(\vec{x}) \right] F(\vec{x},t),
\end{equation}
where the phase-space variables are denoted by $\vec{x} = (\alpha, \alpha^*, \beta, \beta^*, \eta, \eta^*)$, corresponding to $(a, a^\dagger, b, b^\dagger, c, c^\dagger)$.

This representation is mathematically equivalent to an Itô stochastic differential equation~\cite{baxendale2007stochastic}
\begin{equation}
     dx_i = A_i(\vec{x})\,dt + B_{ij}(\vec{x})\,dW_j(t),
     \label{Ito}
\end{equation}
where $dW_j(t)$ are real-valued, independent Wiener processes with $\langle dW_i\, dW_j \rangle = \delta_{ij} dt$. The matrix $B(\vec{x})$ is chosen such that $D(\vec{x}) = B(\vec{x}) B^{T}(\vec{x})$.

From Eqs.~\eqref{FP} and \eqref{Ito}, the corresponding Itô stochastic differential equations can be written as:
\begin{subequations}
\begin{align}
\frac{d\alpha}{dt} &= -i \frac{\omega_a}{4} \alpha - ig \beta \eta - \Gamma \left( |\beta|^2 + \frac{1}{2} \right) \alpha \nonumber \\
&\quad + \sqrt{ \frac{\Gamma\left( |\beta|^2 + \frac{1}{2} \right)}{2} } (dW_1 + i dW_2), \\
\frac{d\beta}{dt} &= i \frac{\omega_a}{4} \beta - ig \alpha \eta^* + \Gamma \left( |\alpha|^2 - \frac{1}{2} \right) \beta \nonumber \\
&\quad + \sqrt{ \frac{\Gamma\left( |\alpha|^2 - \frac{1}{2} \right)}{2} } (dW_3 + i dW_4), \\
\frac{d\eta}{dt} &= -i \omega_c \eta - ig \alpha \beta^* - \kappa \eta
+ \sqrt{ \frac{\kappa}{2} } (dW_5 + i dW_6).
\end{align}
\end{subequations}

By repeating the stochastic simulations over $M \gg 1$ trajectories, the symmetrically ordered expectation values of observables can be estimated by ensemble averaging:
\begin{align}
\langle a^\dagger a (b^\dagger b)_{\rm sym}\rangle &\simeq \frac{1}{M} \sum_{i=1}^{M} |\alpha_i(\beta_i)|^2
\end{align}
where $\alpha_i$ and $\beta_i$ denote the stochastic variables from the $i$-th trajectory.

The TWA captures part of the quantum fluctuations beyond the mean-field approximation by introducing stochastic initial conditions and incorporating random fluctuations during the time evolution. At the same time, it remains computationally efficient for systems with large atom number and enables simulations over long time scales. For comparison, we also examine the mean-field approximation, which can be understood as follows.  In the framework of quantum mechanics, an arbitrary operator can be decomposed into the sum of its expectation value and the fluctuation,  $O = \langle O \rangle + \delta O$ with $\langle \delta O \rangle = 0$. Accordingly, the expectation value of a product of two operators can be expressed as  $\langle AB \rangle = \langle A \rangle \langle B \rangle + \langle \delta A \, \delta B \rangle$. Within the mean-field approximation, the second-order fluctuation term $\langle \delta A \, \delta B \rangle$ is neglected, and only the zeroth-order contribution is retained. Consequently, the expectation value of a product of operators is approximated by the product of their individual expectation values $\langle AB \rangle \approx \langle A \rangle \langle B \rangle$.
Under the mean-field approximation, the time evolution of the operator expectation values obey

\begin{subequations}
\begin{align}
\frac{d\langle S_z\rangle}{dt}&=-ig\langle c\rangle\langle S^+\rangle +ig\langle c^+\rangle\langle S^{-}\rangle\nonumber \\&-2 \Gamma \Big [ \frac{N}{2}(\frac{N}{2}+1)- \langle S_z \rangle^2 + \langle S_z\rangle \Big ],\\
\frac{d\langle S^+\rangle}{dt}&=i\omega_a\langle S^+\rangle -2ig\langle c^+\rangle\langle S_z\rangle-\Gamma\langle S^+\rangle,\\
\frac{d\langle c\rangle}{dt}&=-i\omega_c\langle c \rangle-ig\langle S^-\rangle-\kappa\langle c \rangle.
\end{align}
\end{subequations}

\subsection{Results }
\begin{figure}
\begin{center}
\includegraphics[width=0.8\columnwidth]{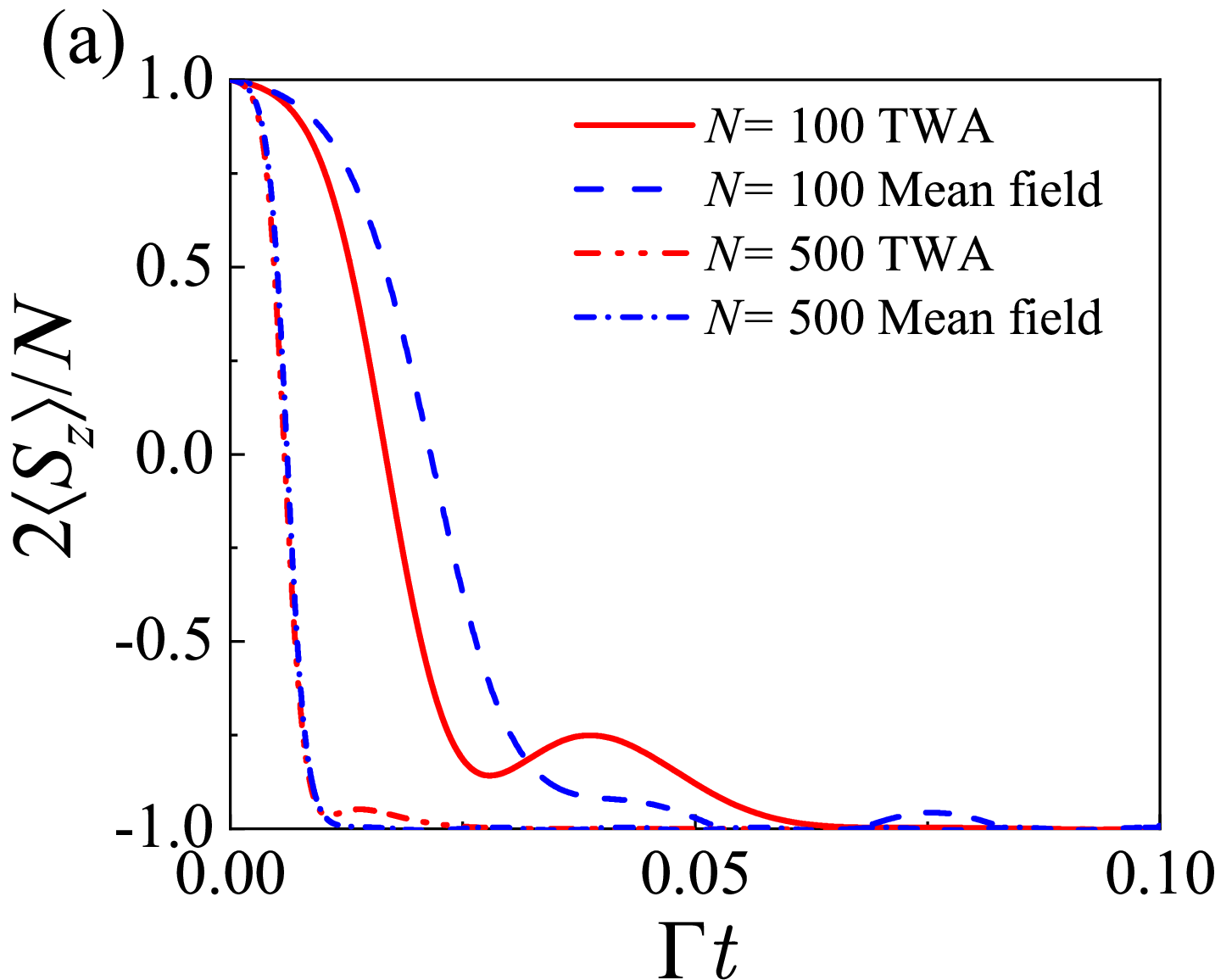}\\[2ex]
\includegraphics[width=0.8\columnwidth]{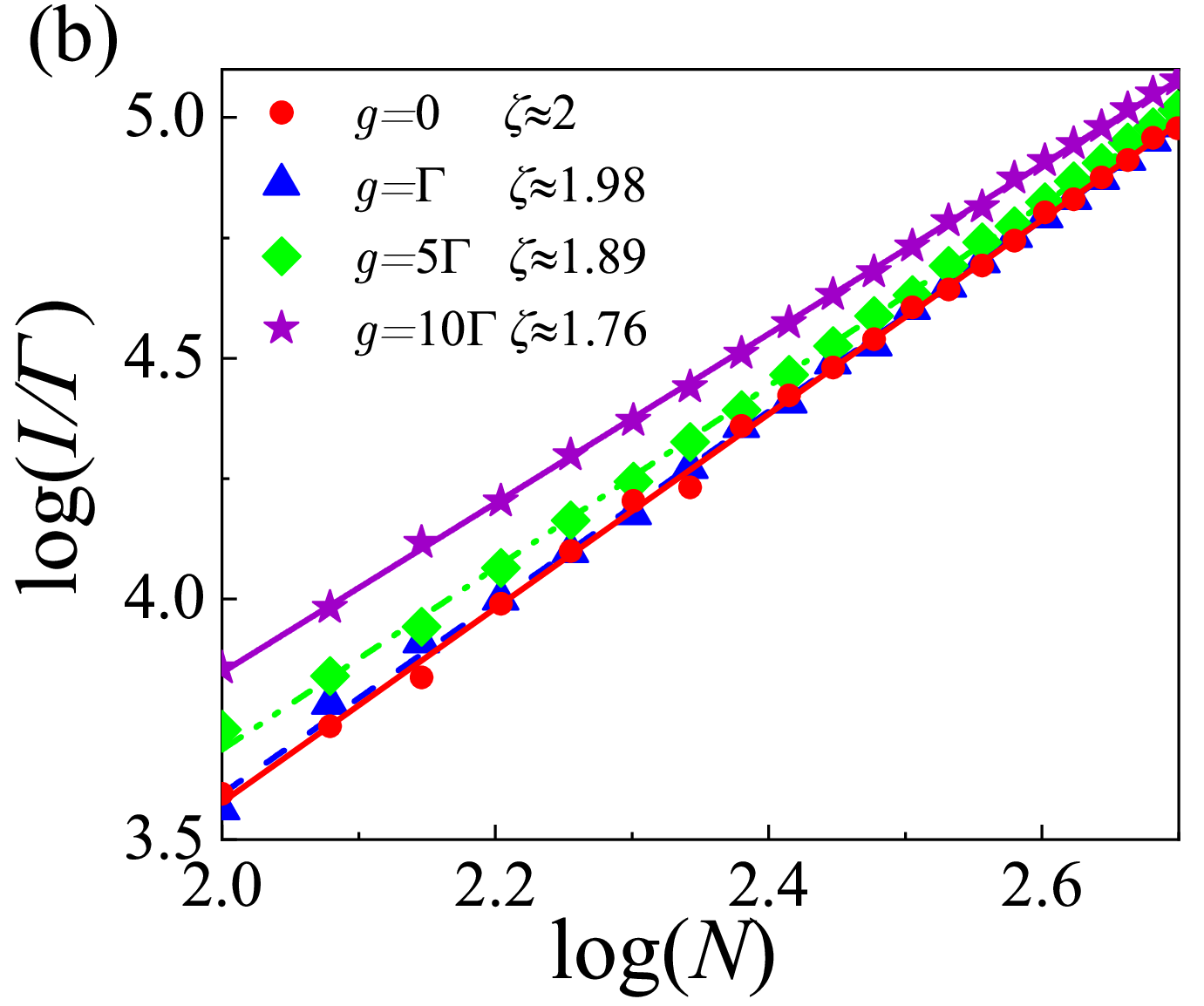}
\end{center}
\caption{\label{FIG.2}(a) Atomic  dynamical evolution $2\langle S_z\rangle/N$ with collective emission, obtained from the TWA and mean-field approximation for atom numbers $N=100$ and $N=500$ with the atom-cavity coupling strength being $g = 10\Gamma$.
(b) Log-log plot of the superradiance strength $I$ versus atom number $N$ for different coupling strengths $g$, while solid lines represent the fitted curves obtained from a power-law fit.  The parameters are chosen as $\omega_a = \omega$ and $\kappa = \Gamma$.
}
\label{collectived}
\end{figure}

 Choosing the initial state $|\psi(0)\rangle=|e_1,e_2,\cdots,e_N;0\rangle$, which represents that all of the atoms are in their excited states while the cavity field is in the vacuum state, we plot the average value of $2\langle S_z\rangle/N$ in Fig~\ref{collectived}(a). As a comparison,  we present both of the results obtained from the TWA and the mean-field approximation for different atom numbers $N$. We observe that the atoms eventually decay to their ground states, that is, $2\langle S_z\rangle/N=-1$, by emitting photons into the cavity field and free space. For a relatively small atom number ($N=100$), the results from TWA and mean-field approximation exhibit significant difference, indicating the importance of quantum fluctuations. In contrast, for a larger ensemble ($N=500$), the two approaches produce good overall results, with small differences, reflecting the suppression of fluctuations in the large-$N$ limit.

In traditional Dicke superradiance without cavity coupling, the collective emission will reach maximum strength at time $t_0$, when the slope of $\langle S_z \rangle$ is maximal and $\langle S_z(t_0) \rangle=0$. The superradiant strength is then defined as $I=d\langle S_z\rangle/dt$ at the moment $t_0$. Due to the collective effect, the traditional superradiant strength yields a quadratic scaling versus the atoms number $N$, i.e., $I \propto N^2$.

To investigate how the coupling to the cavity affects the scaling behavior, we plot the superradiant strength $I$ for various atom-cavity coupling strengths $g$ in Fig.~\ref{collectived}(b). Here, we obtain the discrete points based on the TWA approaches, and then numerically fitting them by the solid lines. The approximately linear dependence of $\log (I)$ on $\log (N)$ indicates that the emission still follows a power-law scaling of the form $I \propto N^\zeta$, where the exponent $\zeta$ corresponds to the slope and depends on the coupling strength $g$. Without coupling to the cavity $g=0$, we reproduce the traditional Dicke superradiance with $\zeta \approx 2$.  For relatively weak coupling, such as $g = \Gamma$, we find $\zeta=1.98$, showing that the weak coupling nearly does not change the Dicke superradiance. However, the scaling exponent gradually decreases with increasing $g$, as shown in Fig.~\ref{FIG.2}(b), where $\zeta\approx1.89$ at $g = 5\Gamma$ and $\zeta \approx 1.76$ at $g = 10\Gamma$. This trend suggests that stronger atom-photon coupling suppresses the ideal superradiant scaling. Nevertheless, for even larger coupling strengths, the system enters a regime where atoms coherently exchange excitations with the cavity field, resulting in damped Rabi oscillations. In this regime, the emission dynamics can no longer be described by simple superradiant scaling and are therefore beyond the scope of the present work.

The above results indicate that the supperradiant behavior of the atomic ensemble is quantitatively modified by weak to moderate coupling between the atoms and the cavity field. In traditional Dicke superradiance without a cavity, all excited atoms emit photons coherently, and the resulting constructive interference leads to a collective emission intensity scaling as $N^2$. When the atoms are weakly coupled to the cavity mode, the influence of the cavity remains minimal, and the system approximately retains the $N^2$ scaling. However, at moderate coupling strengths, the cavity introduces an additional leaky channel. Photons emitted into cavity mode decay, and are not completely shared among the atoms, which degrades the coherence of the collective emission process. This leads to a partial suppression of constructive interference, causing the scaling exponent to be reduced to $\zeta < 2$.

\section{Individual atomic emission}
\label{s3}
\subsection{Model}
\begin{figure}
\begin{center}
\includegraphics[width=0.8\columnwidth]{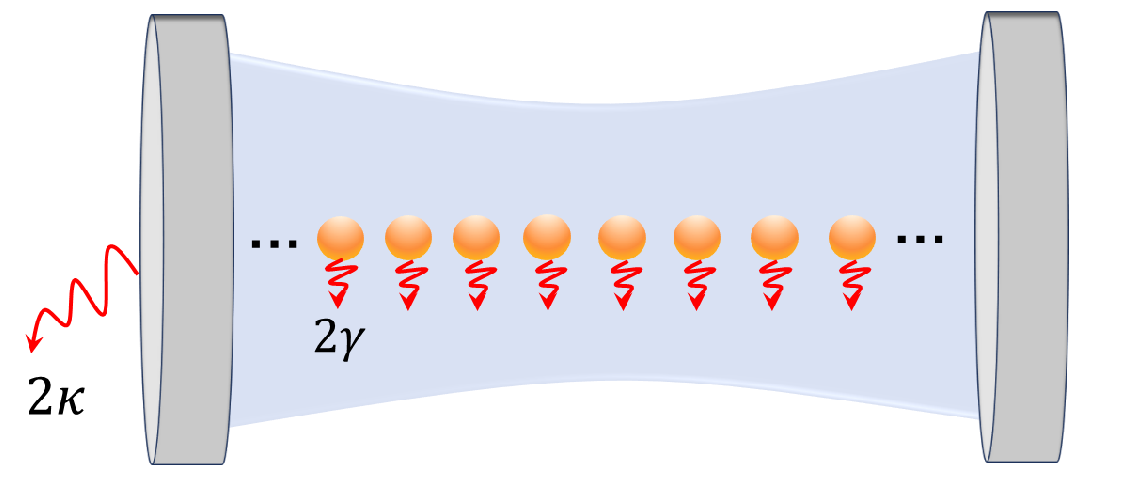}
\end{center}
\caption{\label{FIG.3}Schematic illustration of the individual emission process in a multi-atom cavity QED system. The cavity mode decays at a rate $2\kappa$, and the atoms exhibit individual emission with a decay rate $2\gamma$.}
\end{figure}

In this section, as shown in Fig.~\ref{FIG.3}, we consider an array of $N$ identical two-level atoms that interact with a single-mode cavity,  but each atom is subject to its own spontaneous emission channel. The dynamics of the system is then governed by the master equation
\begin{eqnarray}
{\partial_{t}\rho}=&&-i[H,\rho]+2\kappa(c \rho c^{\dagger}-
\frac{1}{2}\{ c^{\dagger} c, \rho \})\nonumber\\
&&+2\gamma\sum_{i=1}^{N}(\sigma^{i}_{-} \rho \sigma^{i}_{+}-
\frac{1}{2} \{\sigma^{i}_{+} \sigma^{i}_{-}, \rho \}).
\label{masteri}
\end{eqnarray}
where the Hamiltonian $H$ is same as that in Eq.~(\ref{commonH}). The difference between the two schemes lies in the atomic dissipation in Eq.~(\ref{masterq}) and Eq.~(\ref{masteri}). As described in Eq.~(\ref{masterq}), collective atomic emission allows a photon emitted by one of the atoms to be reabsorbed by others. This interatomic interference results in an $N^2$ scaling without coupling to the cavity. However, in the independent atomic emission described by Eq.~(\ref{masteri}), the absence of collective emission leads to a $N$ scaling without cavity coupling.

The master equation for the individual emission case, Eq.~(\ref{masteri}), as in the collective emission scheme, still requires substantial computational resources. However, the TWA, applied in the last section, is not applicable because we cannot rewrite the atomic dissipation using collective spin operators anymore.  Therefore, we employ the DTWA~\cite{schachenmayer2015many,huber2022realistic}  to simulate the atomic dynamics.

Similar to the TWA, the basic idea of the DTWA is to approximate the exact quantum dynamics by evolving $M$ classical noisy trajectories $\vec{s}^i(t)=\langle \vec{\sigma}^i(t)\rangle$, governed by the mean-field dynamical equations. Quantum fluctuations are incorporated through stochastic sampling of the initial conditions from an appropriate probability distribution. In the present superradiance issue with the same initial state as that in the last section, the initial spin configurations are randomly chosen from the four states $(s_x^i,s_y^i,s_z^i)=(\pm1,\pm1,1)$, each occurring with equal probability. The resulting classical variables evolve according to the mean-field Heisenberg equations of motion, $d\langle O\rangle/dt = i \langle[O,H]\rangle$, supplemented by random noise at each time step. The expectation values are then obtained by ensemble averaging over trajectories,
\begin{equation}
    \langle \sigma^i_{j}(t)\rangle \simeq \frac{1}{m}\sum_{i=1}^M {\mathcal S}^i_{j,m}(t),
    \qquad j = x,y,z.
\end{equation}
where ${\mathcal S}^{i}_{j,m}$ represents the value of $s_{j}^{i}$ on the $m$th trajectory.

Starting from the mean-field approximation and incorporating stochastic noise, the It ô stochastic differential equations are
\begin{subequations}
    \begin{align}
    \frac{d s_x^i}{dt} &= -\omega_a s_y^i - g s_z^i (\eta - \eta^*) - \gamma s_x^i - \sqrt{2\gamma}\, s_y^i\, dW_i, \\
    \frac{d s_y^i}{dt} &= \omega_a s_x^i - g s_z^i (\eta + \eta^*) - \gamma s_y^i + \sqrt{2\gamma}\, s_x^i\, dW_i, \\
    \frac{d s_z^i}{dt} &= g \big[ s_y^i (\eta + \eta^*) + s_x^i (\eta - \eta^*) \big] - 2\gamma (s_z^i + 1)
    \nonumber \\&+ \sqrt{2\gamma} (s_z^i + 1) dW_i, \\
    \frac{d \eta}{dt} &= -i \omega_c \eta - \kappa \eta - \frac{i g}{2} \sum_i \left( s_x^i - i s_y^i \right)
   \nonumber \\ & + \sqrt{\frac{\kappa}{2}} (dW_{a1} + i dW_{a2}),
    \end{align}
\end{subequations}
where $\eta = \langle c \rangle$. In the above equations, the cavity mode is treated within the TWA framework, and the noise contribution  to  the atoms is incorporated following the strategy proposed in Ref.~\cite{huber2022realistic}. Here, each $dW_i$ denotes an independent Wiener increment satisfying the It\^{o} rules $\langle dW_i \rangle = 0$ and $\langle dW_i^2 \rangle = dt$, they approximately guarantee the spin-length conservation.

For comparison, we also perform mean-field simulations that neglects inter-operator correlations by factorizing expectation values into products of single-operator averages. The corresponding dynamical equations are
\begin{subequations}
\begin{align}
\frac{d \langle \sigma_z^i \rangle}{dt} &= -2 i g \langle c \rangle \langle \sigma_+^i \rangle + 2 i g \langle c^\dagger \rangle \langle \sigma_-^i \rangle - 2 \gamma (1 + \langle \sigma_z^i \rangle), \\
\frac{d \langle \sigma_+^i \rangle}{dt} &= i \omega_a \langle \sigma_+^i \rangle - i g \langle c^\dagger \rangle \langle \sigma_z^i \rangle - \gamma \langle \sigma_+^i \rangle, \\
\frac{d \langle c \rangle}{dt} &= -i \omega_c \langle c \rangle - i g \langle \sigma_-^i \rangle - \kappa \langle c \rangle.
\end{align}
\end{subequations}

\begin{figure}
\begin{center}
\includegraphics[width=0.8\columnwidth]{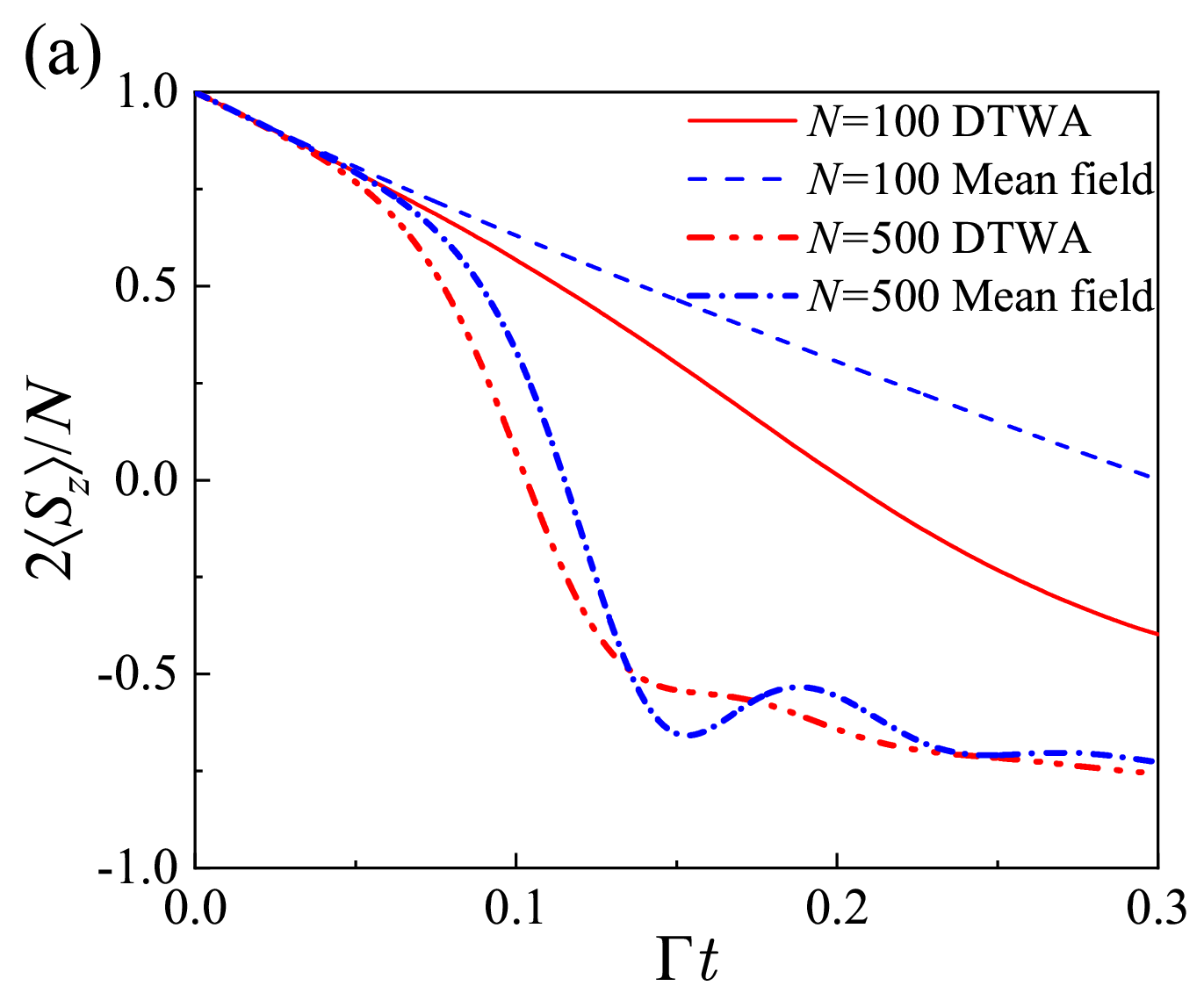}\\[2ex]
\includegraphics[width=0.8\columnwidth]{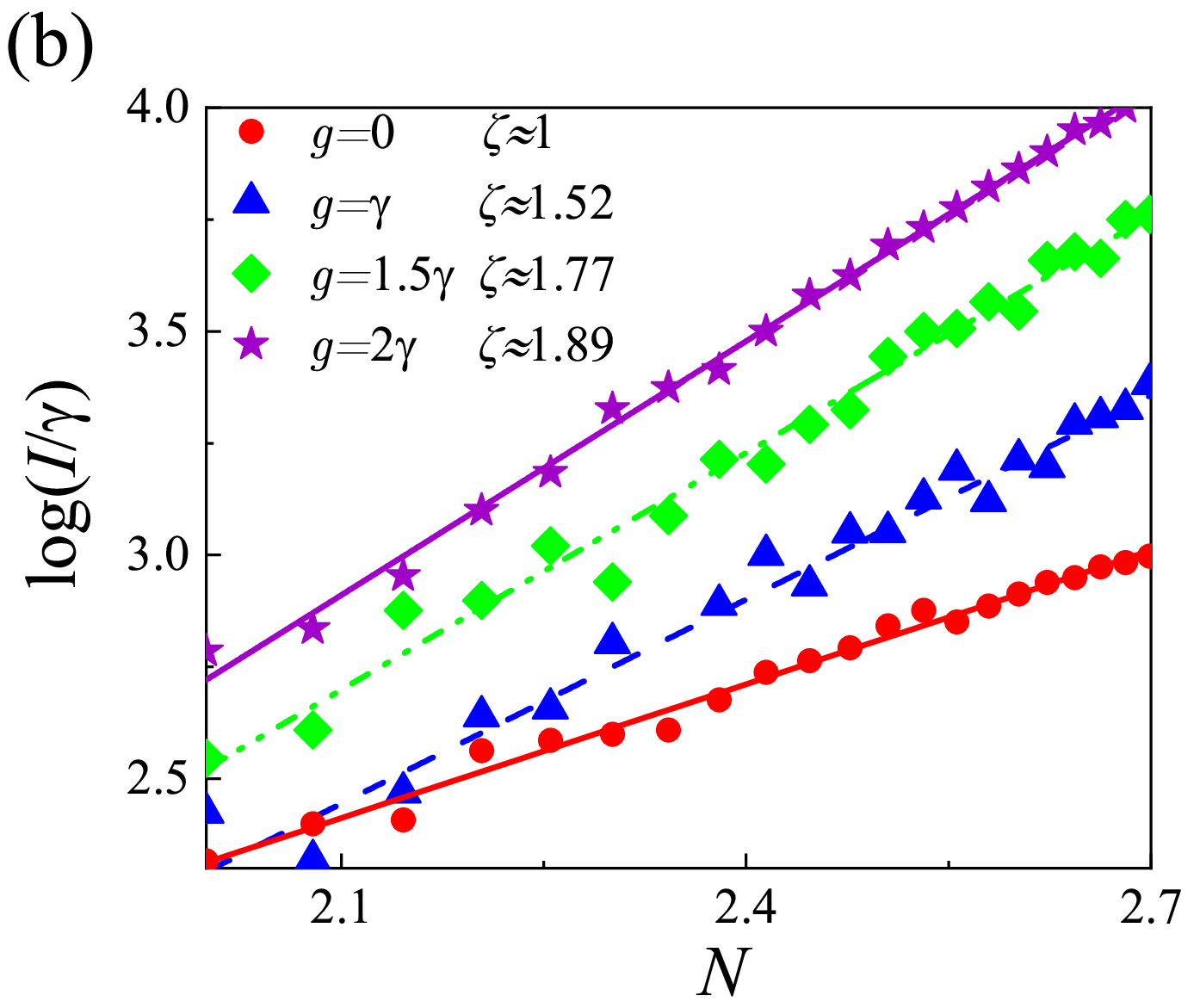}
\end{center}
\caption{\label{FIG.4} (a) Atomic  dynamical evolution $2\langle S_z\rangle/N$ with collective emission, obtained from the TWA and mean-field approximation for atom numbers $N=100$ and $N=500$ with the atom-cavity coupling strength being $g = 2\gamma$.
(b) Log-log plot of the superradiance strength $I$ versus atom number $N$ for different coupling strengths $g$, while solid lines represent the fitted curves obtained from a power-law fit.  The parameters are chosen as $\omega_a = \omega$ and $\kappa = 20\gamma$.
}
\label{idd}
\end{figure}

\subsection{Results}

In Fig.~\ref{idd}(a),  we investigate the time evolution of the average value $2\langle S_z\rangle/N$ for DTWA and the mean-field approximation with atom numbers of $N=100$ and $N=500$ in the case of individual atomic emission. At early times, the results obtained from two approaches agree well for both $N=100$ and $N=500$, indicating that the fluctuation effect, which is not taken into consideration in the mean-field approximation, is negligible. In a longer time evolution, the results from the two approaches begin to deviate from each other. As shown in the Fig.~4(a), the deviation is more obvious for the small atom number of $N=100$ than for the large atom number of $N=500$. This result is consistent with that in the collective emission and shows that the fluctuation plays a more important role in the system with smaller number of atoms.

Although the atoms individually emit in this setup, the leaky cavity provides a common auxiliary channel that induces partial coherence during the radiance process. In Fig.~\ref{idd}(b), we investigate how this channel modifies the emission of atoms by plotting the radiance strength $I$ (which is defined the same as that in the collective emission setup) as a function of the atom number $N$ in a Log-Log scale for different atom-cavity coupling strength $g$. We apply discrete points to represent the results obtained from the DTWA and solid lines to represent the corresponding numerical fitting results. We observe the approximate linear dependence of $\log(I)$ versus $\log(N)$ with the slope $\zeta$, which implies $I\propto N^\zeta$.  Without the coupling to the cavity, i.e., $g=0$, we show that $\zeta \approx 1$. This scaling gradually increases as the atom-cavity coupling strength is enhanced, from $\zeta=1.52$ to $\zeta=1.77$ when the coupling strengths are $g=\gamma$ and $g=1.6\gamma$.  For even stronger coupling strength, e.g., $g = 2 \gamma$, $\zeta$ approaches $1.89$, close to the ideal superradiant scaling of $\zeta = 2$. This result implies that the shared cavity mode acts as an additional channel that can induce collective superradiance even in an otherwise individually emitting atomic ensemble.

\section{Conclusion}
\label{s4}

In conclusion, we have demonstrated that atomic superradiance can be actively controlled by coherently coupling the atoms to a single-mode cavity. For collective atomic emission, the presence of the cavity mode introduces an additional dissipative channel, thereby reducing the radiation scaling from the free-space Dicke limit of $N^2$ to approximately $N^{1.76}$ under moderate coupling conditions. In contrast, for individual atomic emission, the cavity serves as an effective common photonic reservoir, enhancing the emission scaling from the linear $N$ dependence to a superradiant-like behavior of $N^{1.9}$.

The proposed scheme can be readily implemented on contemporary cavity-QED platforms. Experimental advances have already enabled the coupling of several to hundreds of Rydberg atoms to Fabry-P erot~\cite{yan2023superradiant,seubert2025tweezer,hartung2024quantum,wang2025cavity,liu2023realization}, ring~\cite{zhang2024cavity,li2024high}, and fiber cavities~\cite{sunami2025scalable}. In these systems, the atom-photon coupling strength can exceed the cavity decay rate by more than an order of magnitude, making the observation of cavity-controlled superradiance, as discussed in our work, experimentally feasible.

Our results highlight the pivotal role of the cavity in tailoring atomic radiance, offering valuable guidance for experimental efforts aimed at engineering and controlling many-atom dynamics in cavity QED and related platforms.

\begin{acknowledgments}
ZW is supported by the Science and Technology Development Project of Jilin Province (Grant No. 20230101357JC), National Science Foundation of China (Grant No. 12375010) and the Innovation Program for Quantum Science and Technology (No. 2023ZD0300700). RS acknowledges support from the China Scholarship Council (Grant No. 202406620161). AR acknowledges that this publication has emanated from research funded by Taighde \'Eireann-Research Ireland under Grant number 19/FFP/6951.
\end{acknowledgments}

\subsection*{Data availability}
The relevant data is available from the corresponding author upon request.


\begin{thebibliography}{99}

\bibitem{PhysRev.93.99}
R.~H.~Dicke, Coherence in Spontaneous Radiation Processes,
\emph{Phys. Rev.} \textbf{93}, 99--110 (1954).

\bibitem{gross1982superradiance}
M.~Gross and S.~Haroche, Superradiance: An essay on the theory of collective spontaneous emission,
\emph{Phys. Rep.} \textbf{93}, 301--396 (1982).

\bibitem{zhu2024single}
C.~Zhu, S.~C.~Boehme, L.~G.~Feld, A.~Moskalenko, D.~N.~Dirin, R.~F.~Mahrt, T.~St{\"o}ferle, M.~I.~Bodnarchuk, A.~L.~Efros, P.~C.~Sercel \emph{et al.},
Single-photon superradiance in individual caesium lead halide quantum dots,
\emph{Nature} \textbf{626}, 535--541 (2024).

\bibitem{andreev1980collective}
A.~V.~Andreev, V.~I.~Emel'yanov, and Yu.~A.~Il'inskii,
Collective spontaneous emission (Dicke superradiance),
\emph{Sov. Phys. Usp.} \textbf{23}, 493 (1980).

\bibitem{zhang2025unraveling}
X.~H.~H.~Zhang, D.~Malz, and P.~Rabl,
Unraveling superradiance: entanglement and mutual information in collective decay,
\emph{Phys. Rev. Lett.} \textbf{135}, 033602 (2025).

\bibitem{mlynek2014observation}
J.~A.~Mlynek, A.~A.~Abdumalikov, C.~Eichler, and A.~Wallraff,
Observation of Dicke superradiance for two artificial atoms in a cavity with high decay rate,
\emph{Nat. Commun.} \textbf{5}, 5186 (2014).

\bibitem{wang2020supercorrelated}
Z.~Wang, T.~Jaako, P.~Kirton, and P.~Rabl,
Supercorrelated radiance in nonlinear photonic waveguides,
\emph{Phys. Rev. Lett.} \textbf{124}, 213601 (2020).

\bibitem{guo2024phonon}
A.~L.~Guo, L.~T.~Zhu, G.~C.~Guo, Z.~R.~Lin, C.~F.~Li,
 and T.~Tu,
Phonon superradiance with time delays from  collective giant atoms,
\emph{Phys. Rev. A} \textbf{109}, 033711 (2024).

\bibitem{bohnet2012steady}
J.~G.~Bohnet, Z.~Chen, J.~M.~Weiner, D.~Meiser, M.~J.~Holland, and J.~K.~Thompson,
A steady-state superradiant laser with less than one intracavity photon,
\emph{Nature} \textbf{484}, 78--81 (2012).

\bibitem{ferioli2021laser}
G.~Ferioli, A.~Glicenstein, F.~Robicheaux, R.~T.~Sutherland, A.~Browaeys, and I.~Ferrier-Barbut,
Laser-driven superradiant ensembles of two-level atoms near Dicke regime,
\emph{Phys. Rev. Lett.} \textbf{127}, 243602 (2021).

\bibitem{paulisch2019quantum}
V.~Paulisch, M.~Perarnau-Llobet, A.~Gonz{\'a}lez-Tudela, and J.~I.~Cirac,
Quantum metrology with one-dimensional superradiant photonic states,
\emph{Phys. Rev. A} \textbf{99}, 043807 (2019).

\bibitem{koppenhofer2023squeezed}
M.~Koppenh{\"o}fer, P.~Groszkowski, and A.~A.~Clerk,
Squeezed superradiance enables robust entanglement-enhanced metrology even with highly imperfect readout,
\emph{Phys. Rev. Lett.} \textbf{131}, 060802 (2023).

\bibitem{sheremet2023waveguide}
A.~S.~Sheremet, M.~I.~Petrov, I.~V.~Iorsh, A.~V.~Poshakinskiy, and A.~N.~Poddubny,
Waveguide quantum electrodynamics: Collective radiance and photon-photon correlations,
\emph{Rev. Mod. Phys.} \textbf{95}, 015002 (2023).

\bibitem{cardenas2023many}
S.~Cardenas-Lopez, S.~J.~Masson, Z.~Zager, and A.~Asenjo-Garcia,
Many-body superradiance and dynamical mirror symmetry breaking in waveguide QED,
\emph{Phys. Rev. Lett.} \textbf{131}, 033605 (2023).

\bibitem{windt2025effects}
B.~Windt, M.~Bello, D.~Malz, and J.~I.~Cirac,
Effects of retardation on many-body superradiance in chiral waveguide QED,
\emph{Phys. Rev. Lett.} \textbf{134}, 173601 (2025).

\bibitem{mukhopadhyay2024quantum}
D.~Mukhopadhyay and J.-T.~Shen,
Quantum multiphoton Rabi oscillations in waveguide QED,
\emph{New J. Phys.} \textbf{26}, 103026 (2024).

\bibitem{PhysRevLett.110.066802}
H.~Toida, T.~Nakajima, and S.~Komiyama,
Vacuum Rabi Splitting in a Semiconductor Circuit QED System,
\emph{Phys. Rev. Lett.} \textbf{110}, 066802 (2013).

\bibitem{trivedi2019photon}
R.~Trivedi, M.~Radulaski, K.~A.~Fischer, S.~Fan, and J.~Vu{\v{c}}kovi{\'c},
Photon blockade in weakly driven cavity quantum electrodynamics systems with many emitters,
\emph{Phys. Rev. Lett.} \textbf{122}, 243602 (2019).

\bibitem{hamsen2017two}
C.~Hamsen, K.~N.~Tolazzi, T.~Wilk, and G.~Rempe,
Two-photon blockade in an atom-driven cavity QED system,
\emph{Phys. Rev. Lett.} \textbf{118}, 133604 (2017).

\bibitem{kimble1998strong}
H.~J.~Kimble,
Strong interactions of single atoms and photons in cavity QED,
\emph{Phys. Scr.} \textbf{T76}, 127 (1998).

\bibitem{ye1999trapping}
J.~Ye, D.~W.~Vernooy, and H.~J.~Kimble,
Trapping of single atoms in cavity QED,
\emph{Phys. Rev. Lett.} \textbf{83}, 4987 (1999).

\bibitem{boozer2006cooling}
A.~Boozer, A.~Boca, R.~Miller, T.~E.~Northup, and H.~J.~Kimble,
Cooling to the ground state of axial motion for one atom strongly coupled to an optical cavity,
\emph{Phys. Rev. Lett.} \textbf{97}, 083602 (2006).

\bibitem{majumdar2012probing}
 A.~Majumdar, M.~Bajcsy, and J.~Vu{\v{c}}kovi{\'c}, Probing
 the ladder of dressed states and nonclassical light gen
eration in quantum-dot-cavity QED,
\emph{Phys. Rev. A} \textbf{85}, 041801 (2012).

\bibitem{thompson2013coupling}
J.~D.~Thompson, T.~G.~Tiecke, N.~P.~de~Leon, J.~Feist, A.~V.~Akimov, M.~Gullans, A.~S.~Zibrov, V.~Vuleti{\'c}, and M.~D.~Lukin,
Coupling a single trapped atom to a nanoscale optical cavity,
\emph{Science} \textbf{340}, 1202--1205 (2013).

\bibitem{colombe2007strong}
Y.~Colombe, T.~Steinmetz, G.~Dubois, F.~Linke, D.~Hunger, and J.~Reichel,
Strong atom--field coupling for Bose--Einstein condensates in an optical cavity on a chip,
\emph{Nature} \textbf{450}, 272--276 (2007).

\bibitem{liu2023realization}
Y.~Liu, Z.~Wang, P.~Yang, Q.~Wang, Q.~Fan, S.~Guan, G.~Li, P.~Zhang, and T.~Zhang,
Realization of strong coupling between deterministic single-atom arrays and a high-finesse miniature optical cavity,
\emph{Phys. Rev. Lett.} \textbf{130}, 173601 (2023).

\bibitem{yan2023superradiant}
Z.~Yan, J.~Ho, Y.~H.~Lu, S.~J.~Masson, A.~Asenjo-Garcia,
 and D.~M.~Stamper-Kurn, Superradiant and subradiant
 cavity scattering by atom arrays,
\emph{Phys. Rev. Lett.} \textbf{131}, 253603 (2023).

\bibitem{huber2021phase}
J.~Huber, P.~Kirton, and P.~Rabl,
Phase-space methods for simulating the dissipative many-body dynamics of collective spin systems,
\emph{SciPost Phys.} \textbf{10}, 045 (2021).

\bibitem{mink2023collective}
C.~D.~Mink and M.~Fleischhauer,
Collective radiative interactions in the discrete truncated Wigner approximation,
\emph{SciPost Phys.} \textbf{15}, 233 (2023).

\bibitem{huber2022realistic}
J.~Huber, A.~M.~Rey, and P.~Rabl,
Realistic simulations of spin squeezing and cooperative coupling effects in large ensembles of interacting two-level systems,
\emph{Phys. Rev. A} \textbf{105}, 013716 (2022).

\bibitem{khasseh2020discrete}
R.~Khasseh, A.~Russomanno, M.~Schmitt, M.~Heyl, and R.~Fazio,
Discrete truncated Wigner approach to dynamical phase transitions in Ising models after a quantum quench,
\emph{Phys. Rev. B} \textbf{102}, 014303 (2020).

\bibitem{hao2021observation}
L.~Hao, Z.~Bai, J.~Bai, S.~Bai, Y.~Jiao, G.~Huang, J.~Zhao, W.~Li, and S.~Jia,
Observation of blackbody radiation enhanced superradiance in ultracold Rydberg gases,
\emph{New J. Phys.} \textbf{23}, 083017 (2021).

\bibitem{opper2001advanced}
M.~Opper and D.~Saad (eds.),
\emph{Advanced Mean Field Methods: Theory and Practice}
(MIT Press, 2001).

\bibitem{olsen2005phase}
M.~K.~Olsen, L.~I.~Plimak, S.~Rebi{\'c}, and A.~S.~Bradley,
Phase-space analysis of bosonic spontaneous emission,
\emph{Opt. Commun.} \textbf{254}, 271--281 (2005).

\bibitem{ng2011exact}
R.~Ng and E.~S.~S{\o}rensen,
Exact real-time dynamics of quantum spin systems using the positive-$P$ representation,
\emph{J. Phys. A: Math. Theor.} \textbf{44}, 065305 (2011).

\bibitem{baxendale2007stochastic}
P.~H.~Baxendale and S.~V.~Lototsky,
\emph{Stochastic Differential Equations: Theory and Applications}, Vol.~2
(World Scientific, 2007).

\bibitem{schachenmayer2015many}
J.~Schachenmayer, A.~Pikovski, and A.~M.~Rey,
Many-body quantum spin dynamics with Monte Carlo trajectories on a discrete phase space,
\emph{Phys. Rev. X} \textbf{5}, 011022 (2015).

\bibitem{seubert2025tweezer}
M.~Seubert, L.~Hartung, S.~Welte, G.~Rempe, and E.~Distante,
Tweezer-assisted subwavelength positioning of atomic arrays in an optical cavity,
\emph{PRX Quantum} \textbf{6}, 010322 (2025).

\bibitem{hartung2024quantum}
L.~Hartung, M.~Seubert, S.~Welte, E.~Distante, and G.~Rempe,
A quantum-network register assembled with optical tweezers in an optical cavity,
\emph{Science} \textbf{385}, 179--183 (2024).

\bibitem{wang2025cavity}
Z.~Wang, S.~Guan, G.~Teng, P.~Yang, P.~Zhang, G.~Li, and T.~Zhang,
A cavity QED system with defect-free single-atom array strongly coupled to an optical cavity,
\emph{Quantum Frontiers} \textbf{4}, 10 (2025).

\bibitem{zhang2024cavity}
X.~Zhang, Z.~Yu, H.~Zhang, D.~Xiang, and H.~Zhang,
Cavity dark mode mediated by atom array without atomic scattering loss,
\emph{Phys. Rev. Res.} \textbf{6}, L042026 (2024).

\bibitem{li2024high}
Y.~Li and J.~D.~Thompson,
High-rate and high-fidelity modular interconnects between neutral atom quantum processors,
\emph{PRX Quantum} \textbf{5}, 020363 (2024).

\bibitem{sunami2025scalable}
S.~Sunami, S.~Tamiya, R.~Inoue, H.~Yamasaki, and A.~Goban,
Scalable networking of neutral-atom qubits: Nanofiber-based approach for multiprocessor fault-tolerant quantum computers,
\emph{PRX Quantum} \textbf{6}, 010101 (2025).


\end{thebibliography}
\end{document}